\documentclass{cimento}

\usepackage{graphicx}  

\title{The Innovative Design of the PANDA Barrel DIRC}
\author{G.~Schepers\from{ins:1}\thanks{Corresponding author E-Mail g.schepers@gsi.de}\ETC,
A.~Ali\from{ins:1}\from{ins:2},
A.~Belias\from{ins:1},
R.~Dzhygadlo\from{ins:1},
A.~Gerhardt\from{ins:1},
M.~Krebs\from{ins:1}\from{ins:2},
D.~Lehmann\from{ins:1},
K.~Peters\from{ins:1}\from{ins:2},
C.~Schwarz\from{ins:1},
J.~Schwiening\from{ins:1},
M.~Traxler\from{ins:1},
L.~Schmitt\from{ins:3},
M.~B\"{o}hm\from{ins:4},
A.~Lehmann\from{ins:4},
M.~Pfaffinger\from{ins:4},
S.~Stelter\from{ins:4},
F.~Uhlig\from{ins:4},
M.~D\"{u}ren\from{ins:5},
E.~Etzelm\"{u}ller\from{ins:5},
K.~F\"{o}hl\from{ins:5},
A.~Hayrapetyan\from{ins:5},
K.~Kreutzfeld\from{ins:5},
J.~Rieke\from{ins:5},
M.~Schmidt\from{ins:5},
T.~Wasem\from{ins:5},
P.~Achenbach\from{ins:6},
M.~Cardinali\from{ins:6},
M.~Hoek\from{ins:6},
W.~Lauth\from{ins:6},
S.~Schlimme\from{ins:6},
C.~Sfienti\from{ins:6},
M.~Thiel\from{ins:6}}
\instlist{\inst{ins:1} GSI Helmholtzzentrum f\"ur Schwerionenforschung GmbH, Darmstadt, Germany
\inst{ins:2} Goethe University, Frankfurt a.M., Germany
\inst{ins:3} FAIR, Facility for Antiproton and Ion Research in Europe, Darmstadt, Germany
\inst{ins:4} Friedrich Alexander-University of Erlangen-Nuremberg, Erlangen, Germany
\inst{ins:5} II. Physikalisches Institut, Justus Liebig-University of Giessen, Giessen, Germany
\inst{ins:6} Institut f\"{u}r Kernphysik, Johannes Gutenberg-University of Mainz, Mainz, Germany} 
  
\begin{document}

\maketitle

\begin{abstract}
 The Barrel DIRC of the PANDA experiment at FAIR will cleanly separate pions from kaons for the physics program of PANDA. Innovative solutions for key components of the detector sitting in the strong magnetic field of the compact PANDA target spectrometer as well as two reconstruction methods were developed in an extensive prototype program. The technical design and present results from the test beam campaigns at the CERN PS in 2017 and 2018 are discussed.
\end{abstract}
\section{Description}
The fixed target experiment PANDA~\cite{ref:PANDA} at the High Energy Storage Ring (HESR) of the Facility for Antiproton and Ion Research in Europe (FAIR) in Darmstadt will offer unique opportunities to solve fundamental questions of hadron physics by using a cooled high-intensity antiproton beam with beam momenta from 1.5 to 15 GeV/$c$. Two fast and compact ring imaging Cherenkov detectors using the DIRC (Detection of Internally Reflected Cherenkov light) technology (the Barrel DIRC and the Endcap Disc DIRC) are foreseen to provide excellent charged particle identification (PID) in the PANDA target spectrometer.
In those detectors Cherenkov photons are produced in thin radiator bars or plates, which are highly polished and to a high precision rectangular shaped to guide effectively the photons via internal reflection to the photon sensors while conserving the angle of the Cherenkov cone. Simulations show that in the compact PANDA detector a downsized version of the barrel shaped BaBar DIRC~\cite{ref:BaBar}, which used conventional photomultiplier tubes covering the surface of a large expansion volume filled with purified water, would reach the performance goals of PANDA. However, since the readout volume of the final PANDA Barrel DIRC has to fit in the tight space of the solenoid with a magnetic field of about one Tesla at the photon sensors new concepts concerning the size and material of the expansion volume as well as the type of photon sensors had to be employed. Detailed Geant4 simulations~\cite{ref:tdr}~\cite{ref:tdrJINST} were performed to optimize the design for performance and cost. 
\begin{figure}[t]
\begin{center}
\includegraphics[scale=0.18,angle=-90]{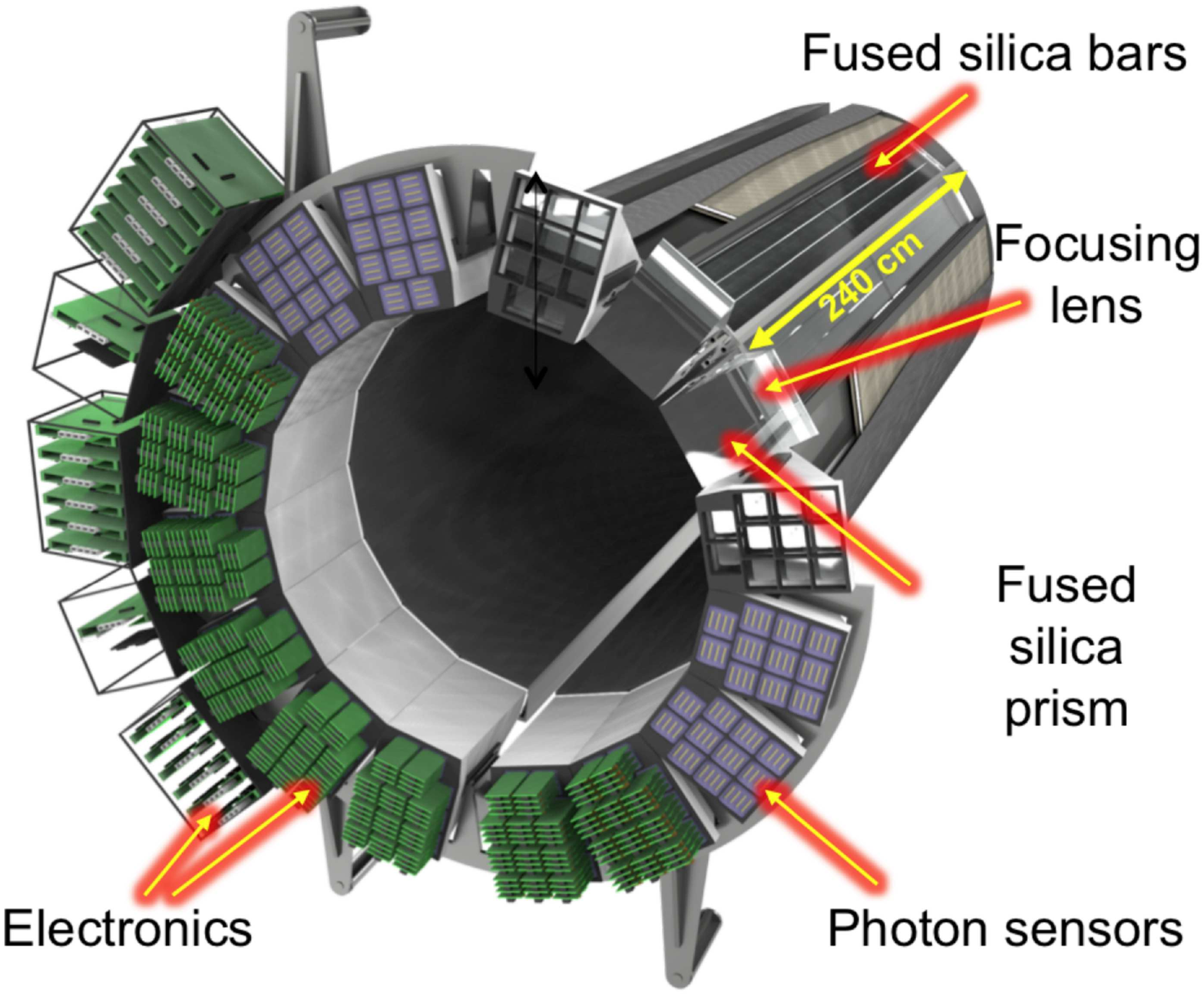} 
\caption{The baseline design of the PANDA Barrel DIRC \label{fig:barreldirc}} 
\end{center}
\end{figure}
The PANDA Barrel DIRC~\cite{ref:tdr}~\cite{ref:tdrJINST} (see fig.~\ref{fig:barreldirc}) covering the polar angle range from 22 deg to 144 deg is required to separate pions from kaons for momenta up to 3.5 GeV/c by 3 standard deviations (s.d.). The baseline design consists of 16 optically isolated sectors, each comprising a barbox with three bars, flat mirrors (on the downstream side) and focusing lenses as well as a compact fused silica prism as expansion volume, and 11 Microchannel-Plate Photomultiplier Tubes (MCP-PMTs) as photon sensors. An ultra fast FPGA based readout system is connected to the MCP-PMTs. Each bars consists of synthetic fused silica, has a length of 2400 mm and a cross section of 17 mm x 53 mm, and is built to tight optical and mechanical specifications to preserve the photon angle during many internal reflections and to optimize the light transport efficiency.
The modular setup of the PANDA Barrel DIRC allows to separate the readout volume part from the barboxes to provide access to the inner detectors. 
\section{Three layer spherical lens and photon sensors}
\begin{figure}[t]
\begin{center}
\includegraphics[scale=0.19,angle=-90]{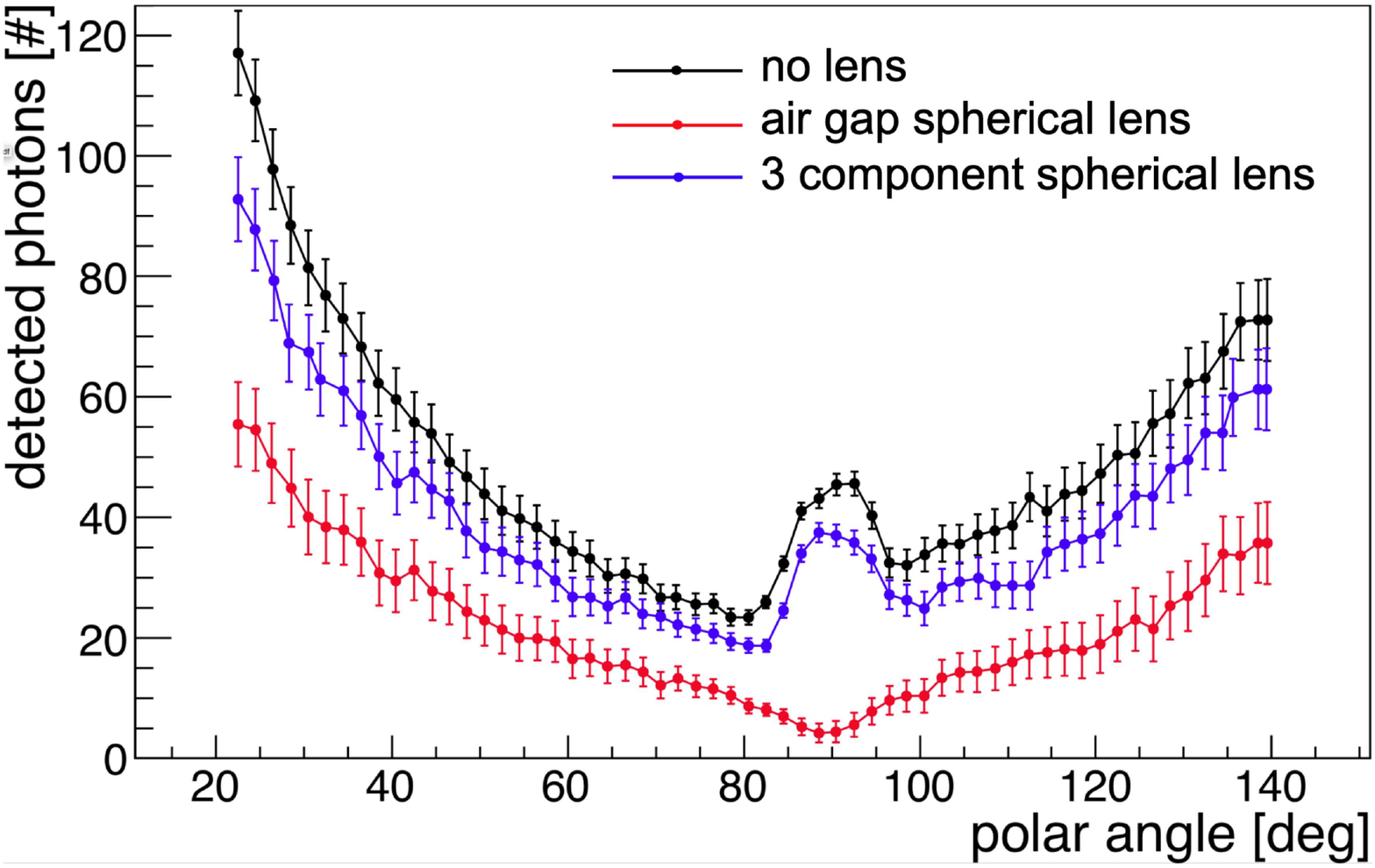}
\hspace{8mm}
\includegraphics[scale=0.20,angle=-90]{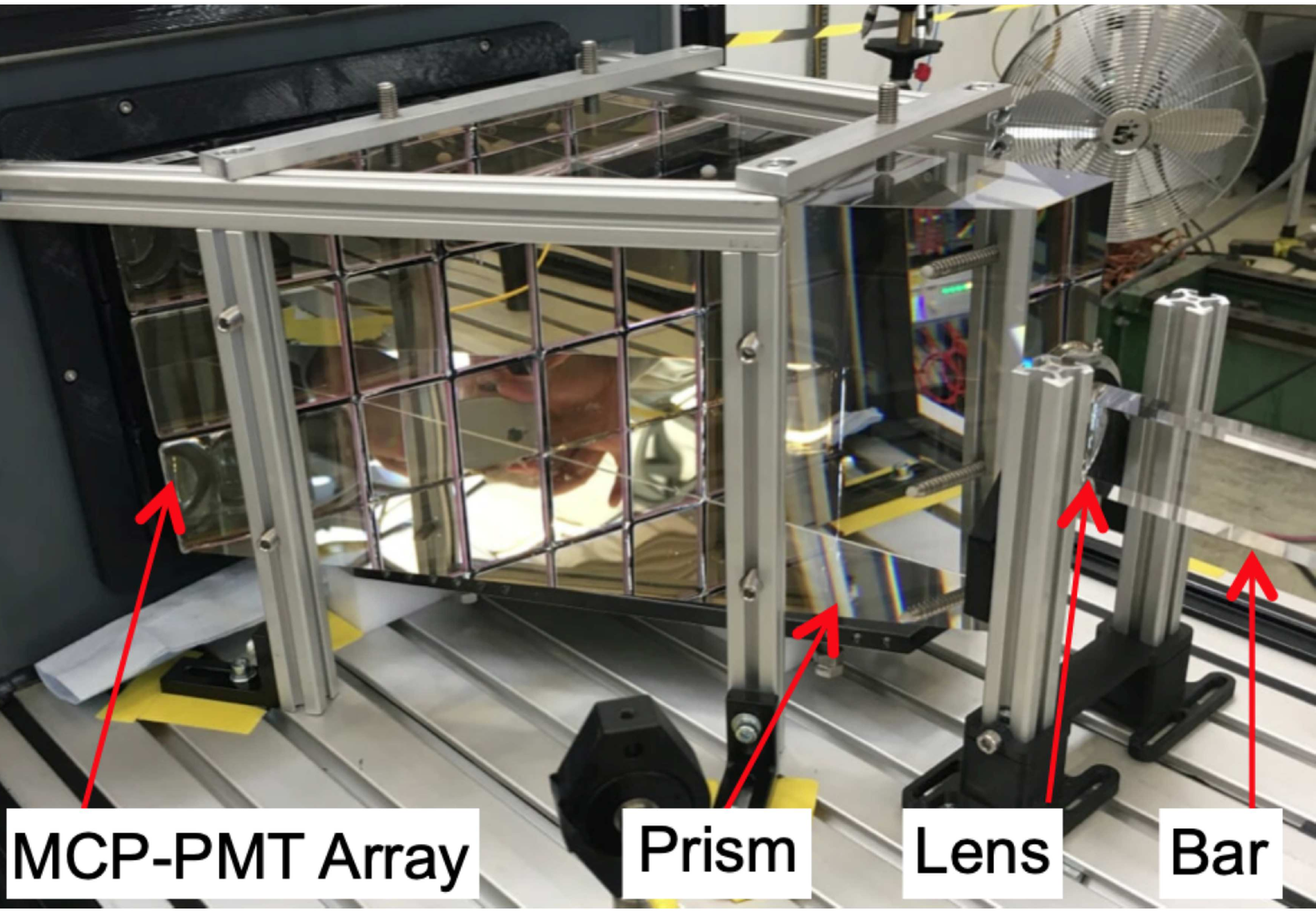}
\caption{The three layer spherical lens avoids holes in the acceptance (left). \label{fig:lens} Photo in the optical box of the prototype at CERN/2017 (right). \label{fig:setup}}
\end{center}
\end{figure}

Due to the compact expansion volume having a length of only 300 mm the need of focusing was investigated. With a conventional lens added between the radiator bar and the expansion volume (which are both from synthetic fused silica) no flat focal plane was achieved and a loss of photons with steep angles was noted compared to the direct coupling of the bar to the prism (see fig.~\ref{fig:lens}, left, ~\cite{ref:lens}). Moreover a large hole in the DIRC acceptance was generated due to reflections at the air gap between the lens and the prism. 
In cooperation with the optical industries a multi layer spherical lens system was designed to efficiently focus the Cherenkov light on a flat image plane on the back wall of the prism. In this innovative design the refraction between the higher-refractive index material lanthanum crown glass (LaK33B, refraction index n about 1.786) used as middle layer in the 3-layer spherical lens and fused silica (n about 1.437) was employed avoiding the adverse air gap. As shown in the figure the number of detected photons is increased and the hole in the acceptance was closed. In fig.~\ref{fig:lens}, right, the prototype setup of the beamtime at CERN 2017 with the lens is shown.

The environmental condition of the PANDA experiment with a high magnetic field of about one Tesla at the position of the photon sensors as well as the high time precision, low noise and single photon detection capacity needed leave MCP-PMTs as the only candidates for the photon detection. However, only recently a few producers could deliver MCP-PMTs that could withstand the expected high rate of the PANDA experiment of $2\cdot10^7$ reactions per second and the calculated accumulated anode charge of about $5 C/cm^2$ of 10 years of PANDA operation.
In close collaboration with the photon sensor industries several countermeasures against the rapid decrease of the quantum efficiency (QE) were studied. Recent lifetime-enhanced MCP-PMTs~\cite{ref:sensors} with MCPs treated with the atomic layer deposition technique exceed the requirements for the PANDA DIRC counters and keep high QE values over the lifetime of PANDA. 
\section{Validation of PID performance in particle beams at CERN 2017}
Several beam times at GSI and CERN were performed to validate the figures of merit of the prototypes achieved in detailed Geant simulations. The number of photons needs to be larger than 20 and the single photon resolution (SPR) less than 10 mrad over all polar angles covered by the Barrel DIRC.
\begin{figure}[t]
\begin{center}
\includegraphics[scale=0.14,angle=-90]{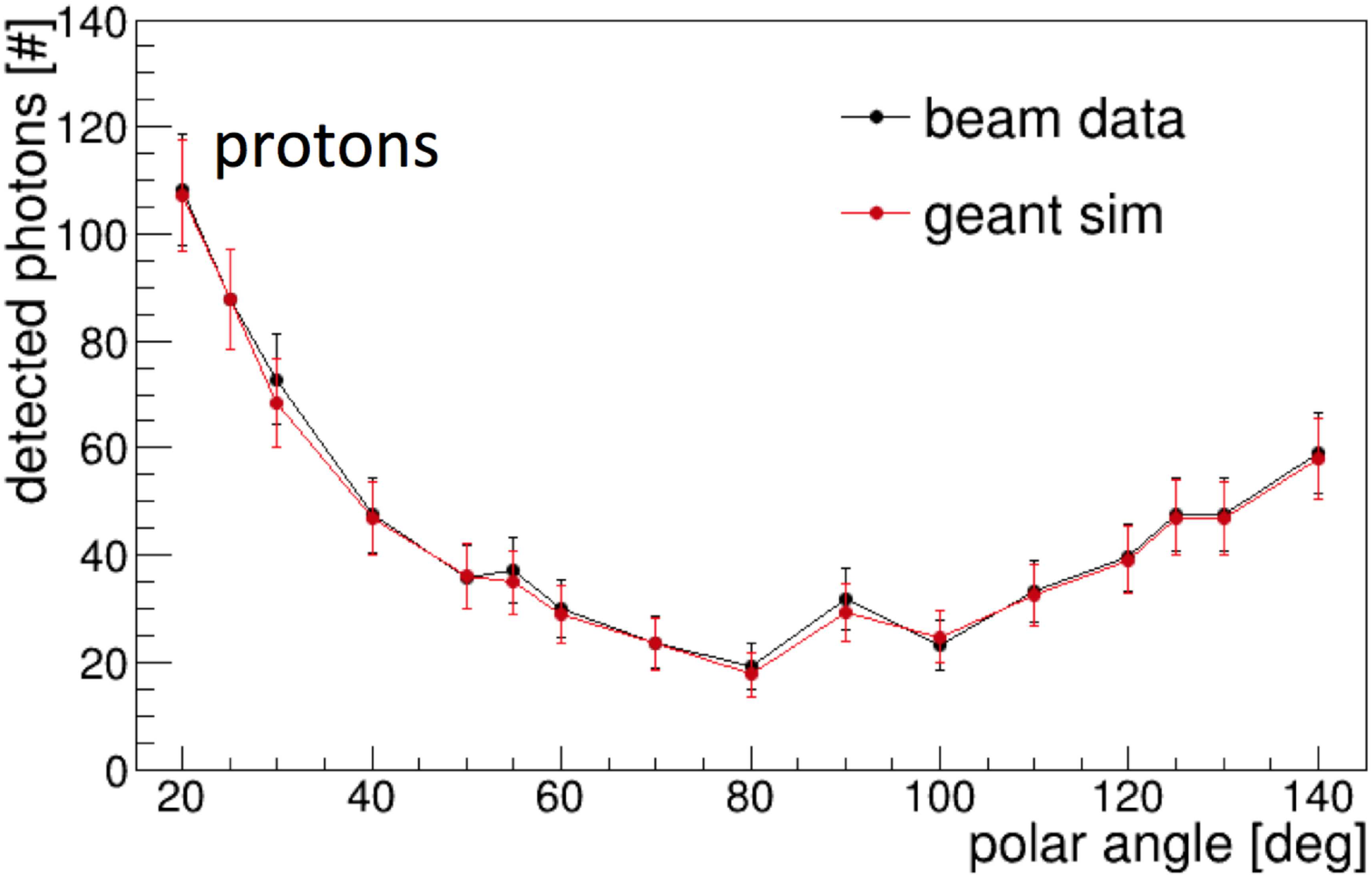} 
\includegraphics[scale=0.14,angle=-90]{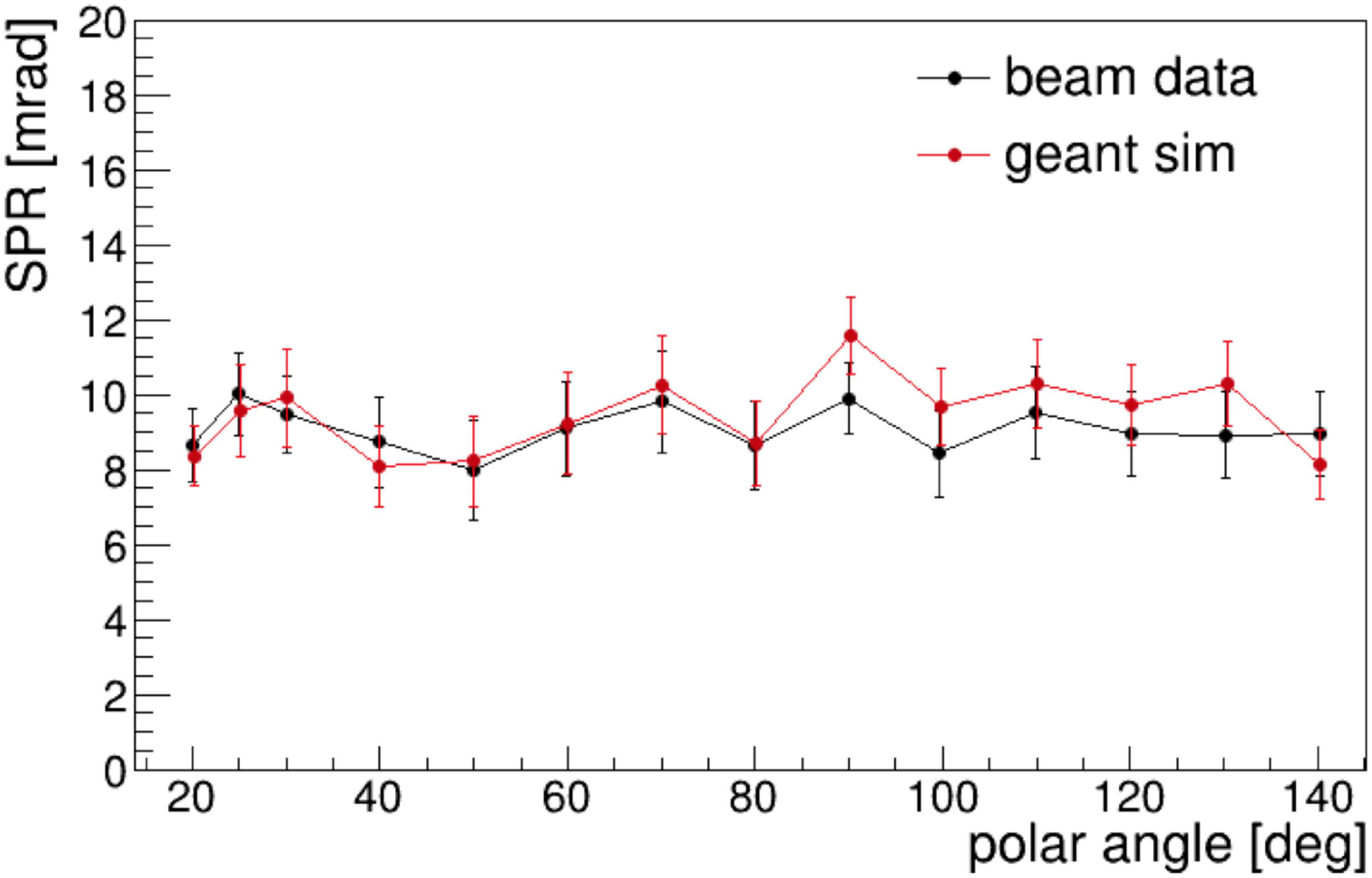}
\includegraphics[scale=0.16,angle=-90]{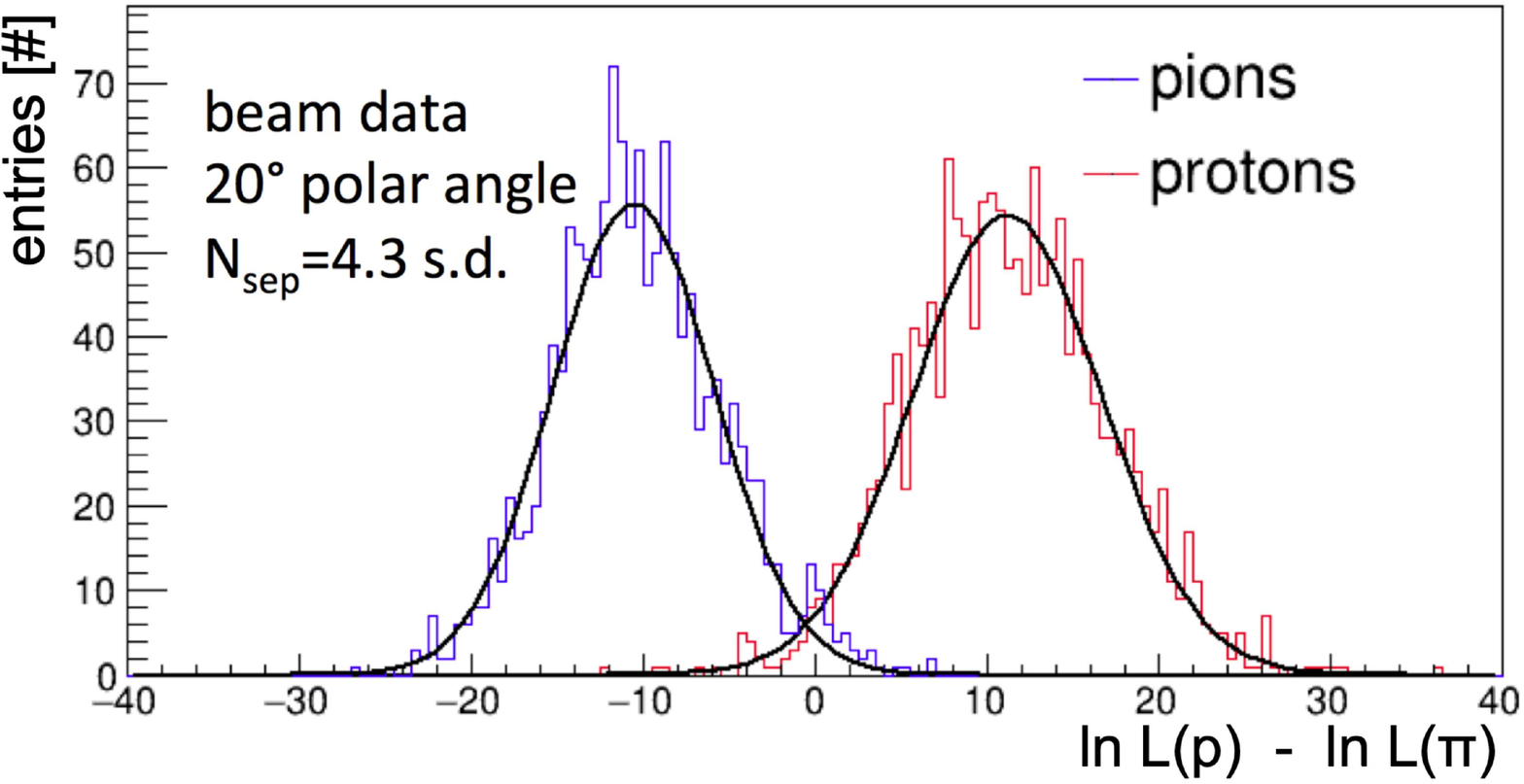} 
\caption{Results of the beamtime CERN 2017. Very good agreement between simulation and prototype data (CERN/2017) for the number of detected photons (left), single photon resolution (middle) and separation power for the prototype setup at CERN 2017 (right) \label{fig:PID}. \label{fig:figuresofmerit}}
\end{center}
\end{figure}
In 2017 a prototype (see fig.~\ref{fig:setup}) consisting of a single radiator of 35 mm width with a mirror at one end (not on the photo) and a lens and a prism covered with 12 MCP-PMTs on the other end was evaluated. The whole prototype could be turned in polar angle as well as in azimuthal angle with respect to the beam line of the CERN T9 area of the PS complex. A set of triggers and a hodoscope as well as two time of flight counters with a flight path of about 29 meters in between and a time precision of 130 ps (sigma) were used. With this external PID information the protons and pions from the beam could be cleanly tagged. Since the Cherenkov angle difference of pions and protons at beam momenta of 7 GeV/$c$ are nearly the same as for kaons and pions at 3.5 GeV/$c$ the PID performance of the prototype could be directly tested.
Two complementary reconstruction algorithms were developed. With the "BaBar-like" geometrical reconstruction method, primarily based on photon spatial coordinates, the photon yield (fig.~\ref{fig:figuresofmerit}, left) and the SPR (fig.~\ref{fig:figuresofmerit}, middle) of the beam data were compared to the Geant simulation. A very good agreement was achieved meaning that the envisaged goals were reached and even exceeded.
The PID-performance of the setup was obtained with the "Belle II TOP-like"~\cite{ref:Belle} time-imaging method, which emphasizes the precise measurement of the photon propagation time to each individual pixel. The probability density functions for the propagation time of the Cherenkov photons from pions and protons were produced from tagged beam data.
The average timing precision of the setup at CERN was 250 ps. From the log-likelihood-difference for kaons and pions including all pixels hit per event (see fig.~\ref{fig:figuresofmerit}, right) the separation power of 4.3 s.d. was achieved for the most challenging forward region of 20 deg. 
Preliminary results from the beam time of August/September 2018 show, that even a reduced number of only 8 photo sensors per expansion volume still reaches the required figures of merit. Extrapolating the results from the 2017 prototype to the fully equipped PANDA Barrel DIRC with a time precision of 100 ps suggests an achievable pions-kaon separation of up to 6.6 s.d. at 25 deg.  

\section{Summary}
The PANDA Barrel DIRC provides excellent PID in the compact space of the PANDA target spectrometer employing focusing and ultrafast timing. The performance of the innovative and cost-optimized setup was evaluated in particle beams at CERN.
The tender process of the most time consuming components of the detector starts in the end of summer 2018 to be ready for installation and commissioning within the PANDA detector in 2023/24.

 
\acknowledgments
This work was supported by HGS-HIRe, HIC for FAIR and BNL
71 eRD14. We thank the CERN staff for the opportunity
to use the beam facilities and for their on-site support.

\end{document}